\documentclass[conference]{IEEEtran}

\IEEEoverridecommandlockouts

\usepackage{cite}
\usepackage{amsmath,amssymb,amsfonts}
\usepackage{algorithmic}
\usepackage{graphicx}
\usepackage{textcomp}
\usepackage{xcolor}
\def\BibTeX{{\rm B\kern-.05em{\sc i\kern-.025em b}\kern-.08em
    T\kern-.1667em\lower.7ex\hbox{E}\kern-.125emX}}

\usepackage{tabularx,booktabs}
\newcolumntype{C}{>{\centering\arraybackslash}X} % centered version of "X" type
\usepackage{caption}
\usepackage{subcaption}

\usepackage{listings}
\usepackage[utf8]{inputenc}
\usepackage{tikz}
\usetikzlibrary{arrows, shapes, positioning}
\begin{document}
\title{A New Software Framework for Traffic Engineering: Path Cardinality and the Effect of Multipath on Residual Capacity\\
\thanks
}

\author{\IEEEauthorblockN{Mohammed Salman and Bin Wang}
\IEEEauthorblockA{\textit{Department of Computer Science and Engineering} \\
\textit{Wright State University, Dayton, OH 45435 USA}\\
\{salman.6, bin.wang\}@wright.edu}
}

\maketitle

\begin{abstract}
In this paper, we present a new traffic engineering (TE) software framework to analyze, configure, and optimize (with the aid of a linear programming solver) a network for service provisioning.
The developed software tool is based on our new data-driven traffic engineering approach that analyzes a large volume of network configuration data generated given the user input. By analyzing the data, one can then make efficient decisions later when designing a traffic engineering solution.

We focus on three well-known traffic engineering objective functions: minimum cost routing (MCR), load balancing (LB), and average delay (AD). With this new tool, one can answer numerous traffic engineering questions. For example, what are the differences among the three objective functions? What is the impact of an objective function on link utilization? How many candidate paths are enough to achieve optimality or near-optimality with respect to a specific objective. This new software tool allows us to conveniently perform various experiments and visualize the results for performance analysis. As case studies, this paper presents examples that answer the questions for two traffic engineering problems: (1) how many paths are required to obtain a solution that is within a few percent from the optimal solution and whether that number is fixed for any network size? (2) how the choice of single-path/multi-path routing affects the load in the network? For the first problem, it turns out that the number of paths needed to achieve optimality increases as the number of links in the network increases.
%
%WWW: I mean based on this software framework, we can collect data and then analyze these data. And based on that, one can make decisions later when designing a complete TE system.if it is not accurate, we can remove it.
%
For the second problem, we show how multi-path routing impacts the residual capacity in the network. As it turns out, multi-path routing achieves better performance than single-path routing in terms of the residual capacity. The power and utilities of this tool can be harnessed to solve other challenging traffic engineering problems efficiently.

%WWW: I have added this sentence above "As it turns out, multipath routing has better advantage over singlepath routing in terms of the residual capacity."

\end{abstract}


\section{Introduction}
With the development of software defined networking (SDN), a new type of traffic engineering system has been the focus of some recent research (e.g., \cite{B4}  \cite{YATES} \cite{SWAN} \cite{Heorhiadi} \cite{SIMPLE} and many others). The basis of all of these approaches is a centralized TE model. The output of a TE system includes a routing scheme that produces a set of candidate paths and the split ratio of traffic among these paths.
%In the language of SDN, these correspond to the control plane and the data plane, respectively. Thus, it appears that
To answer some of the questions related to this centralized TE model is an important step before developing an SDN TE application.
Most of these questions center on TE objective functions, path cardinality, and multi-path/single-path routing. This work aims to develop a traffic engineering software framework that focuses on these research questions using a data-driven approach. We then apply the software tools developed to a few case studies to provide insight to service provisioning with traffic engineering.

The software framework has been developed with the following objectives in mind:

\begin{itemize}
  \item \textbf{Repeatability:} The tool should be able to repeat some of the TE studies conducted by other researchers easily.

  \item \textbf{Abstraction:} The TE framework should hide the complexity of expressing network optimization problems given a variety of parameters and network configurations as formulating and solving a linear program (LP) or a mixed-integer linear program (MILP) is a nontrivial task.
%that as they expose a very low-level interface.

  \item \textbf{Extendability:} The tool should readily allow a new feature to be added fast as all the modules are developed and implemented by following the best practices of software engineering.
\end{itemize}

The framework comes with some additional benefits:

\begin{itemize}
\item \textbf{Making observations:} New observations can be made by studying the effect of a set of parameters that have not been studied in the past without having to dive into the optimization and/or mathematical details.
\item \textbf{Exploring different aspects of the problem:} For example, single-path and multi-path routing may have the similar effect on the optimal value for large networks. However, as we will see later, each type of approach may utilize network link resources differently.
\end{itemize}

This new TE framework and tool will enable students and researchers to express new network traffic engineering optimization problems and setting various parameters and requirements conveniently. The data produced are then stored in a dataset that is ready for analysis and visualization, and for developing solutions to traffic engineering. The effectiveness of our approach have been demonstrated with case studies. The power and utilities of this tool can be harnessed to solve other challenging traffic engineering problems efficiently. The developed software tool, including the data, will be freely available for researchers to repeat published studies \footnote{https://github.com/MohammedSalman/TE-Viz}.

%WWW: url for the software

%WWW: I have added the url as a footnote.

The rest of the paper is organized as follows. In Section \ref{Traffic-Engineering-Models}, we discuss the three popular objective functions that are implemented in this framework. In Section \ref{TE-Overview}, we give an overview about the software framework. This includes how to configure a network, the implemented traffic matrix models, and some discussion on the design choices made in the framework (e.g., path-based and link-based formulation). Section \ref{TE-Framework-Design} presents the design challenges and implementation. The results and the discussion of two case studies on path cardinality and multi-path/single-path routing are depicted in Section \ref{Data-Generation-Examples}. Finally, the paper concludes with future work and discussion in Section \ref{Conclusions}.

\section{Traffic Engineering Models} \label{Traffic-Engineering-Models}
We consider a multi-commodity network flow problem (MCNFP) and a directed communication network $G=(N, L)$ where $N$ is the set of nodes and $L$ is the set of links in graph $G$. Each link $l$ has a capacity $c \geq 0$ for each $l \in L$. A set of demands is denoted as $D$, with traffic volume $h_{d}$, $d \in D$. We consider the case when all nodes in the network send and receive traffic to/from all nodes in the network and only when the system is feasible to accommodate the demands. All the TE models will be expressed using the path-based (a.k.a. link-path) formulation \cite{b1}.

%WWW: path-based or link-path? or do we have a choice? what do you mean in the above parentheses?

%WWW: path-based has another name for it which is (link-path) and it is based on paths (path-centric) while the other type called Node-Link formulation which is (edge-centric).These are discussed in the book of Deepankar Medhi I sent you before. The book is also listed in the references.
A TE model takes the network topology and traffic demands as input and finds how much traffic must flow on path $p$ as output for each source-destination (SD) pair. In case of multi-path routing, the goal is to find the optimal set of flow variables $x = (x_{dp} : d \in D, p \in P_{d})$ where $P_{d}$ is the set of paths for demand $d$. In the case of single-path, the goal is to find one single-path for each SD pair that must be used to achieve a specific optimality.
The goal of TE is to traffic engineer paths to ensure good performance in terms of an optimization objective (e.g., minimizing the delay) and efficient use of network resources (e.g., load balancing). Currently, in our framework, the implemented objective functions  include load balancing (LB), average delay (AD), and minimum cost routing (MCR).

In the following, we briefly introduce these objective functions and problem formulations.

\subsection{Minimum Cost Routing (MCR)}
The MCR objective is to minimize the routing cost, and the problem formulation is given as follows \cite{b1}:

\begin{small}
\begin{subequations}\label{MCRequation:main}
\begin{align}
& \text{min}  && F(x) = \sum_{d \in D_{}}^{} \sum_{p \in P_d}^{} \xi _{dp} x_{dp}     &   & \tag{\ref{MCRequation:main}} \\
& \text{s.t.} && \sum_{p \in P_d}^{}x_{dp} = h_d,                                     & d & \in D \label{MCRequation:d}  \\
&             && \sum_{d \in D~}^{} \sum_{p \in P_d}^{} \delta_{dpl} x_{dp} \leq c_l, & l & \in L \label{MCRequation:c}
\end{align}
\end{subequations}
\end{small}

\begin{small}
\noindent where:\\
\begin{tabular}{l p{7cm}}
$\xi$ & the total cost associated with each path\\~\\
$x_{dp}$ & the flow on path p for demand $d$ \\
$h_{d}$ & the demand volume for demand $d$ \\
$ c_{l}$ & the capacity for link $l$ \\
$ P_{d}$ & the number of candidate paths for demand $d$ \\
$ \delta _{dpl}$ & link-path indicator, either 0 or 1. 1 if path $p$ for demand $d$ uses link $l$; 0 otherwise.\\~\\
\end{tabular}
\end{small}

The constraint \eqref{MCRequation:d} ensures that traffic demands are met and is the same as constraint \eqref{LBequation:d}. The constraint \eqref{MCRequation:c}, the capacity constraint, ensures that the load $\sum _{d \in D~ }^{} \sum _{p \in P_{d} }^{} \delta _{dpl} x_{dp} $ does not exceed the capacity $c_{l}$ of link $l$.

\subsection{Load Balance (LB)}

The LB objective is also known as the minimization of the maximum link utilization, or the Wozencraft objective \cite{wozencraft}, or simply as congestion minimization. The objective is to minimize the load on the most congested link. The formulation \cite{b1} for the LB problem can be written as:

\begin{small}
\begin{subequations}\label{LBequation:main}
\begin{align}
& \text{min}  && F(x) = r     &   & \tag{\ref{LBequation:main}} \\
& \text{s.t.} && \sum _{p \in P_{d} }^{}x_{dp}= h_{d},                                     & d & \in D \label{LBequation:d}  \\
&             && \sum_{d \in D~}^{} \sum_{p \in P_d}^{} \delta_{dpl} x_{dp} \leq c_l r, & l & \in L \label{LBequation:c}
\end{align}
\end{subequations}
\end{small}

% \noindent where:\\
% \begin{tabular}{l p{7cm}}
% $x_{dp}$ & the flow on path p for demand $d$ \\
% $h_{d}$ & the demand volume for demand $d$ \\
% $ c_{l}$ & the capacity for link $l$ \\
% $ P_{d}$ & the number of candidate paths for demand $d$ \\
% $ \delta _{dpl}$ & link-path indicator, either 0 or 1. 1 if path $p$ for demand $d$ uses link $l$; 0 otherwise.\\~\\
% \end{tabular}

The first constraint \eqref{LBequation:d}, the demand constraint, is to make sure that all demands are satisfied over some paths. The second constraint \eqref{LBequation:c} is to ensure that the load does not exceed the capacity of link $l$ multiplied by $r$ that we are trying to minimize. The value of $r$, after solving the linear program, should be less than or equal to 1. Otherwise, the load on some links exceeds the capacity.
% Note that the solution to the above linear program is always feasible because there is no explicit capacity constraints to limit the load.
%To prevent this from happening, we have to add the same set of capacity constraints from the MCR problem \eqref{MCRequation:c}.
%
%WWW: isn't the constraint 2b?
%
%WWW: that is right. I removed these sentences by commenting them out.
%
The linear program formulation above is the final form of the problem and the original problem is non-linear. To see how the problem can be converted to the current form, the reader may refer to Chapter 4 of \cite{b1}.

\subsection{Average Delay (AD)}
\begin{figure}[t]
  \centering
   \includegraphics[width=65mm]{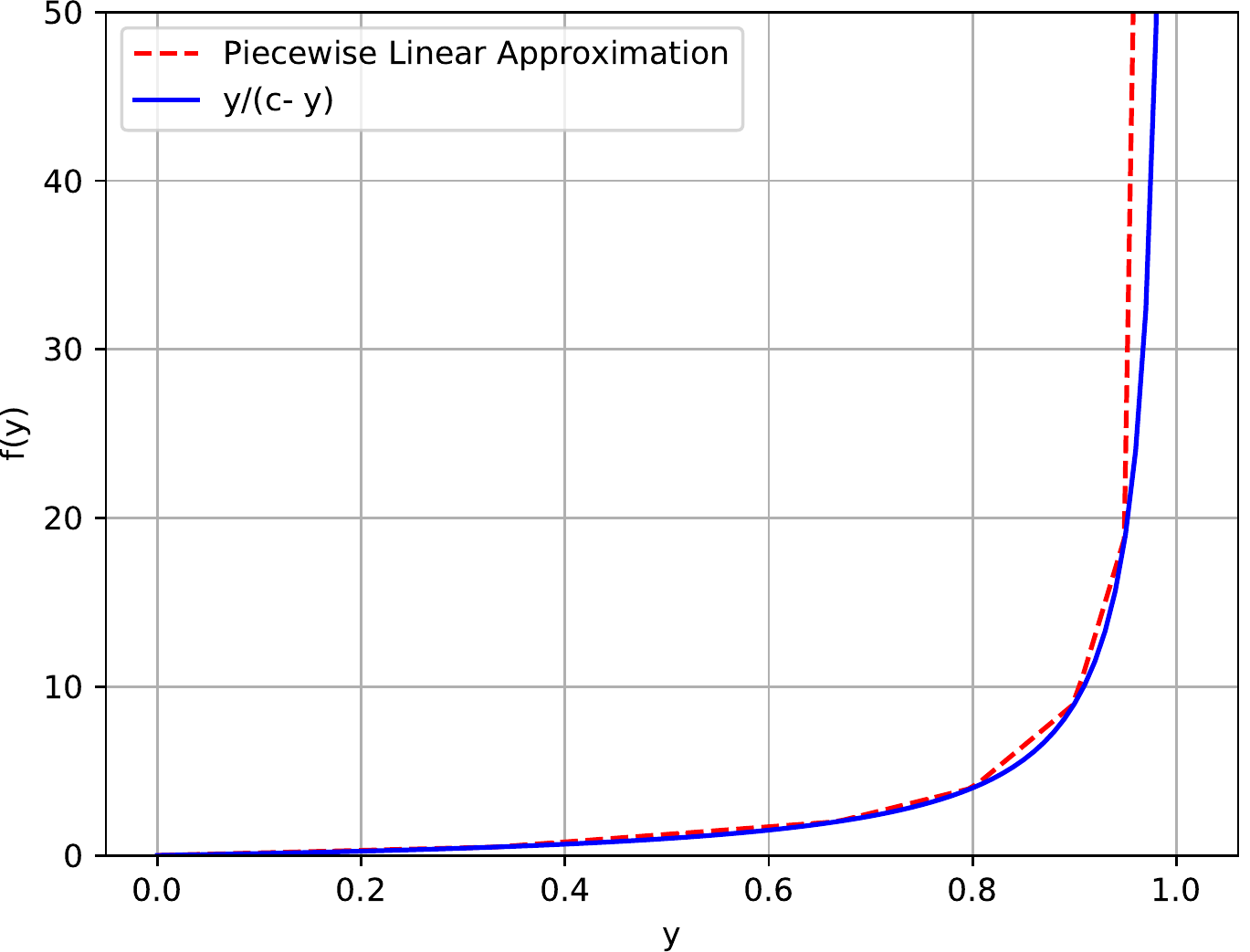}
  \caption{Piecewise linear approximation of the delay function.}
  \label{pla}
\end{figure}

For this objective function, the delay of a link in a network can be modeled with the equation $y/(c-y)$ where $y$ is the load on the link and $c$ is the capacity of the link (see Figure~\ref{pla}). Similar to the LB objective, the original AD problem is non-linear and cannot be formulated directly as a linear program. Thus, this delay function is approximated with a piecewise linear approximation (PLA) function \eqref{ADequation} (see the dotted line in Figure~\ref{pla}).

\begin{small}
\begin{equation}\label{ADequation}
g(z) = \begin{cases}
(3/2)z &\text{for $1 \leq z < 1/3$}\\
(9/2)z-1 &\text{for $1/3 \leq z < 2/3$}\\
15z-8 &\text{for $2/3 \leq z < 4/5$}\\
50z-36 &\text{for $4/5 \leq z < 9/10$}\\
200z-171 &\text{for $9/10 \leq z < 19/20$}\\
4000z-3781 &\text{for $z \geq 19/20$}
\end{cases}
\end{equation}
\end{small}

The linear program for this AD problem is given as follows \cite{b1}:

\begin{small}
\begin{subequations}\label{ADLP:main}
\begin{align}
& \text{min}  && F= \sum _{l=1}^{L}\frac{r_{l}}{c_{l}}     &   & \tag{\ref{ADLP:main}} \\
& \text{s.t.} && \sum _{p=1}^{P_{d}}x_{dp}=h_{d},                                    \quad d= 1, 2,...,D   \\
&             && \sum _{d=1}^{D} \sum _{p=1}^{P_{d}} \delta _{dpl} x_{dp}= y_{l},  \quad l= 1, 2,..., L  \\
&             && r_{l} \geq \frac{3}{2} y_{l}, 	   \quad l= 1, 2,..., L  \\
&			  && r_{l} \geq \frac{9}{2} y_{l}- c_{l}, \quad	 l= 1, 2,..., L  \\
&			  && r_{l} \geq 15 y_{l}- 8c_{l}, 		\quad l= 1, 2,..., L  \\
&			  && r_{l} \geq 50 y_{l}- 36c_{l},		\quad l= 1, 2,..., L  \\
& 			  && r_{l} \geq 200 y_{l}- 171c_{l}, 	\quad l= 1, 2,..., L  \\
&			  && r_{l} \geq 4000 y_{l}- 3781c_{l},	\quad l= 1, 2,..., L  \\
& 		   	  && x_{dp} \geq 0, 					\quad p=1,2,..., P_{k}, d=1, 2, ..., D \\
& 			  && y_{l} \geq 0, 						\quad l= 1, 2,..., L
\end{align}
\end{subequations}
\end{small}

The above linear program formulation is claimed in \cite{b1} to be much more accurate than the approximation of Fortz et al. \cite{Bernard} \cite{X.Liu}.

%WWW: can you explain why this formulation is more accurate?

%WWW: There is no explanations in the original source about why it is more accurate, but I think by "more accurate" they mean that their approximation (the six equations/lines) fits the original curve better than the one approximated by Fortz et al.

\section{TE Framework Overview} \label{TE-Overview}
Our goal in this work is to develop a software tool to allow the user to better understand the effects of different sets of parameters on TE optimization objectives while avoiding the nuances of low-level optimization formulations, making the testing and tuning of TE more accessible to researchers and students.

\begin{figure}[t]
\begin{tikzpicture}[node distance=4mm, >=latex',
 block/.style = {draw, rectangle, minimum height=10mm, minimum width=30mm,align=center},
rblock/.style = {draw, rectangle, rounded corners=0.5em, minimum width=8mm} ]
    \node [rblock]                      (start)     {Start};
    \node [block, right=of start]       (igf)   {Input Configuration\\ File};
    \node [block, right=of igf]     (fati)  {Formulate All the\\ Instances};
    \node [block, below=of fati]     (sati)  {Solve All\\ the Instances};
    \node [block, below=of igf]     (satritd)  {Store All the Results\\ in the Dataset};
    \node [rblock, left=of satritd]                  (end)       {End};

  \path[draw,->] (start)      edge (igf)
                   (igf)    edge (fati)
                   (fati)   edge (sati)
                   (sati)       edge (satritd)
                   (satritd)   edge (end)
                    ;

\end{tikzpicture}
  \caption{The four main steps of the developed framework.}
  \label{sequenceDiagram}
\end{figure}
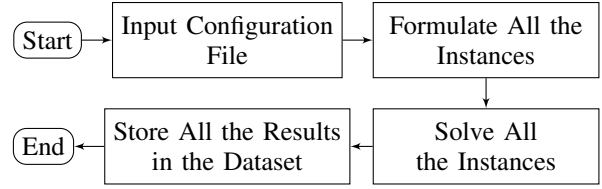

Figure \ref{sequenceDiagram} shows the four main steps of the developed TE software framework.
The framework takes network and other parameters as an input configuration file (e.g., number of nodes, number of links, traffic matrix models, optimization requirements, and so on). The framework will find different sets of combinations of these input parameter, formulate a proper optimization problem, and treat each one (with a different set of input parameters) as an instance that is ready to be solved to obtain an optimal solution (with the aid of an LP solver).
The optimization requirement (i.e., TE objective) is also provided as input to the framework.
%WWW: do not understand the above sentence
%WWWW: do not understand the following sentence
%The tool traffic engineer routes to find the optimal flow on these routes based on the optimization requirement (TE objective) which is also must be provided as input to the framework.
%WWW: do not understand the above sentence
%WWW: By "optimization requirement" I mean the objective function that the user specify in the input file. If it is vague, we can remove it.
The output from the framework is a dataset that has all the information for each instance of the problem solved by the LP solver. The dataset obtained is ready for analysis and visualization. The size of the generated data depends on the number of problem instances contained in the input configuration file. The framework is open for extension. Therefore, additional features can be added to generate needed information for TE to be included in the dataset for analysis and visualization.

In this TE framework, the user can define a set of parameters as inputs to establish a network. These parameters include the number of nodes $n$ and the number of links $l$\footnote{Anywhere in the paper, when we refer to \textit{number of links} we mean \textit{number of link pairs}. If any two nodes are connected, they are connected in both directions.}.
For example, Listing \ref{python_example} specifies a set of network parameters in Python. This TE framework creates a number of topologies by taking the combination of N and L and distributing links between nodes using the Erdős–Rényi model. If the resulting topology is disconnected, the framework will continue to create new ones until the topology is connected or stops after attempting a specified number of times and failing to produce a valid network topology. The Erdős–Rényi model can find a connected topology immediately if $l >> n$. Other configuration information that needs to be fed to the system is shown in Listing \ref{python_example}. The information includes capacity type, capacity set, weight setting, demand traffic matrix model(s), the objective function(s), number of candidate paths, and routing strategies.
The example below is an input configuration file that creates 9600 instances of TE problem, each configured differently due to different links distribution and/or different parameter settings. The number 9600 is calculated from $|N| \times |L| \times Nu\_of\_TMs\_Per\_TOPO \times Nu\_of\_TOPOs\_Per\_N\_L \times |TM\_TYPES| \times |OBJECTIVES| \times |CANDIDATE\_PATHS| \times |ROUTING\_STRATEGIES|$. The parameter  $Nu\_of\_TOPOs\_Per\_N\_L$ is the number of topologies for each pair of $N$ and $L$, and the parameter  $Nu\_of\_TMs\_Per\_TOPO$ is the number of traffic matrices per topology.

\lstset{language=Python}
\lstset{frame=lines}

\lstset{caption={An input configuration file.}}
\lstset{label={lst:code_direct}}
\lstset{basicstyle=\footnotesize}
\begin{lstlisting} [label=python_example]
N = {30, 40}
L = {80, 100}
Nu_of_TMs_Per_N_L = 10
Nu_of_TOPOs_Per_N_L = 10
capacity_type = {'EDGE_BETWEENNESS'}
capacity_set = {30, 35, 40}
weight_setting = {'INV_CAP'}
tm_types = {'BIMODAL', 'GRAVITY'}
Network_Load = [0.07]
objectives = {'LB', 'AD', 'MCR'}
candidate_paths = {3, 7}
routing_strategies = {'MULTIPATH', 'SINGLEPATH'}
\end{lstlisting}

\subsection{Configuring Link Capacity}
Currently, the software tool implements the \textbf{‘edge\_betweenness’} \cite{Lorenzo} method of configuring link capacity. This model takes the set of capacity (CAPACITY\_SET) as input and distributes them based on metrics that capture the importance of links and nodes. In \cite{Lorenzo}, three models were suggested for this task and these are edge betweenness centrality, degree centrality gravity, and communicability centrality gravity. The one that we use is edge betweenness centrality. Edge betweenness centrality is an important metric that measures the “criticality” of a link or a node \cite{Tizghadam}. For example (see Figure \ref{4NodeTopo}), the link (2, 3) is configured the highest capacity which is 40 because this link is the only link that connects node 2 to all other nodes in the network. Likewise, for the other nodes, one can see why the capacities are assigned as shown in the figure.

\subsection{Link Weights}
\begin{figure}
  \centering
   \includegraphics[width=0.30\textwidth, angle=0]{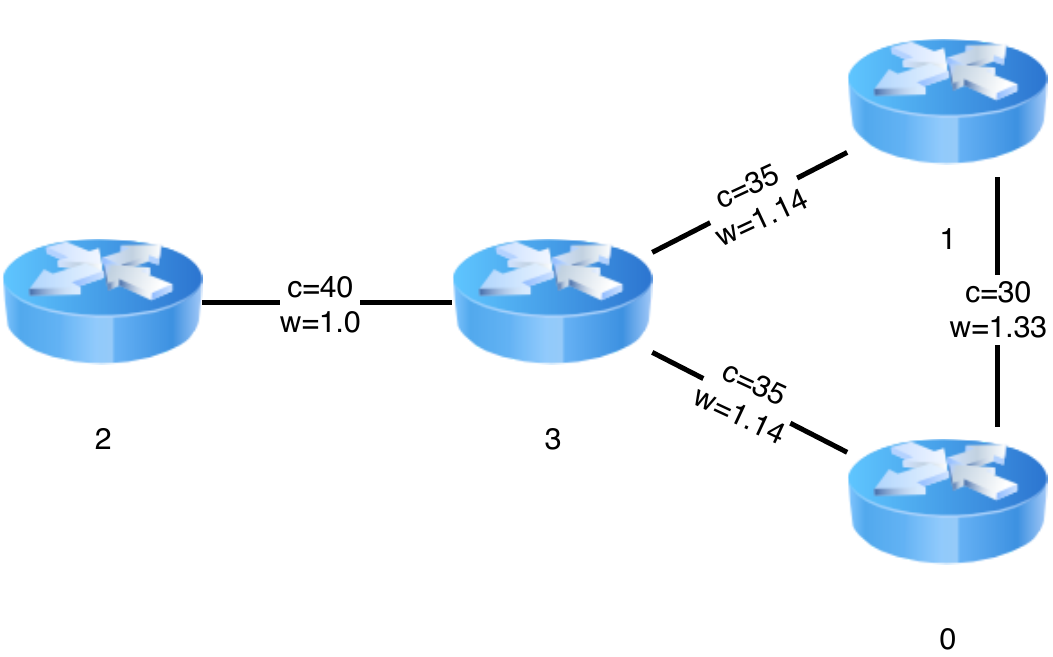}
  \caption{A four-node topology shows how the weights are configured proportionally to the inverse of the link capacity.}
  \label{4NodeTopo}
\end{figure}

In all our experiments and as suggested by Cisco, links are set proportionally to the inverse of link capacity by setting the parameter (WEIGHT\_SETTING) to ‘INV\_CAP’. For example (see Figure \ref{4NodeTopo}), these weights are used to calculate the shortest paths for any pair of nodes in the network.

\subsection{Path-based vs. Link-based Formulation}
We believe that expressing the network optimization problems using path-based (a.k.a Link-Path) formulation is more beneficial than casting them using link-based (a.k.a Node-Link) formulation (despite the fact that we study the link load in some parts of our study) for the following reasons: First, our main purpose of this work is to study the behavior of the network in terms of paths that are selected to attain performance optimality or near-optimality. Second, it is normally easier for application developers and network operators to think in terms of paths instead of links. Finally, in an SDN switch, it is easy to model the routing table based on traffic carried on paths that traverse that switch. Furthermore, key policy requirements for a spectrum of applications are expressed in terms of paths. For example, traffic engineering and service chaining \cite{Heorhiadi}. All the TE objectives can be expressed using the path-based formulation. Bi et. al. \cite{Yingjie} mentioned in their study that path cardinality constraint (i.e., the number of paths allowed) greatly affects the attainable performance of a given routing scheme and the actual implementation complexity. For all the reasons above, we use path-based over link-based formulation. However, we do not use the full set of available paths between each ingress-egress nodes. In fact, to reach optimality or near-optimality, we only need no more than 5 paths as we will show later. The use of a full set of available paths will introduce two efficiency problems (since the number of paths grows exponentially as the size of the network increases). First, calculating all of these paths may take a long time. Second, using all the available paths will only increase the number of variables in the LP formulation, hence, increasing the size and the complexity of the optimization problem.

\subsection{Multi-path vs. Single-path}
Although the LP and MILP problems in networking design optimization can be mitigated by omitting unnecessary long paths, the problems are still hard to solve in case that the traffic flows are unsplittable and only one single-path is permitted for each traffic demand. The single-path constraint introduces discreteness (discrete constraints) into the optimization problem. An additional binary variables are introduced to the problem to decide whether the path per traffic demand should be used or not. Solving the single-path problem for small topologies should not require a significant amount of time, especially with the high-performance computation power that we have today, but when we start increasing the number of nodes and links, the problem becomes intractable.

\subsection{ISP Topologies vs. Arbitrary Topologies}
In our opinion, it is important to consider arbitrary topologies rather than being limited to existing well-known ISP topologies. One reason is that we may want to study the characteristics of TE no matter what the topology is. The use of available ISP topologies may restrict us from fully understanding our TE techniques. On the other hand, in \cite{X.Liu} the authors considered only topologies that may not represent the current ISP topologies.

\subsection{Traffic Matrix Models}
In all our experiments, each node sends to and receives from all other nodes in the network. Three models are implemented to generate traffic matrixes (i.e., Nucci Heuristic \cite{Nucci}, Gravity \cite{gravity}, and Bimodal \cite{bimodal}). These models can be chosen using the parameter TM\_TYPES. We consider using static traffic matrices only and exclude dynamic traffic matrices from our experiments. In a static traffic matrix, the traffic volumes are collected at a single point in time. The generation of Nucci Heuristic is based on the work proposed by Nucci et al. \cite{Nucci}, which comprises of three steps to generate the static traffic matrix. The reader can refer to (\cite{Lorenzo}, \cite{Nucci}) to learn how the traffic matrix is generated. For all the three models, one common parameter ($Max_u$) needs to be set to represent the target maximum link utilization used for scaling traffic volumes. The bimodal model \cite{bimodal} assumes that only a few ingress-egress pairs have very large flows and then assigns demands for these pairs uniformly. The gravity model \cite{gravity} is based on capacity-based heuristic (also used in \cite{sculpte}, \cite{cohen_applegate}) which assumes that each demand for each node is proportional to the combined capacity of connecting links.

The TE framework will then translate all of the inputs above into low-level constraints using Gurobi, the mathematical programming solver \cite{gurobi}. Finally, the framework stores the results in a dataset. Table \ref{table1} shows the information the framework may produce.

%\lipsum[1] % filler text
\begin{table*}
 \caption{Description of the information produced by the framework}
\label{table1}
\begin{tabularx}{\textwidth}{@{}l*{10}{l}c@{}}
\toprule
Parameter     & Type  & Description \\
\midrule
n                           	& discrete      & Number of nodes used          \\
l                         	    & discrete      & Number of links used         \\
Avg\_nodal\_degree              & discrete      & Average nodal degree           \\
Network\_load                   & discrete      & Network loads (traffic scales) used in the traffic matrix         \\
TM\_TYPE                    	& discrete      & Traffic matrix generator models (‘lognormal’, ‘gravity’, ‘bimodal’)           \\
Obj\_type                  		         & discrete      & Objective function used          \\
Obj\_val                    		       & continuous      & Optimal or near-optimal value of the objective function           \\
k                           						& discrete      & Number of candidate paths used           \\
Routing\_strategy                           & discrete      & Multipath and/or Single-path           \\
Residual\_cap                  	         & continuous      & Percentage of available capacity after solving the problem instance           \\
Links\_utilization\_and\_residual     & dictionary data structure      & Nested data structure for storing information about all the links in each instance           \\
Flow\_agg\_per\_path                     	      & dictionary data structure      & Nested data structure for storing information about all routes in each instance           \\

%\addlinespace

\bottomrule
\end{tabularx}
\end{table*}
%\lipsum[2-15] % more filler text

\section{TE Framework Design} \label{TE-Framework-Design}
\begin{figure}
  \centering
   \includegraphics[width=0.40\textwidth, angle=0]{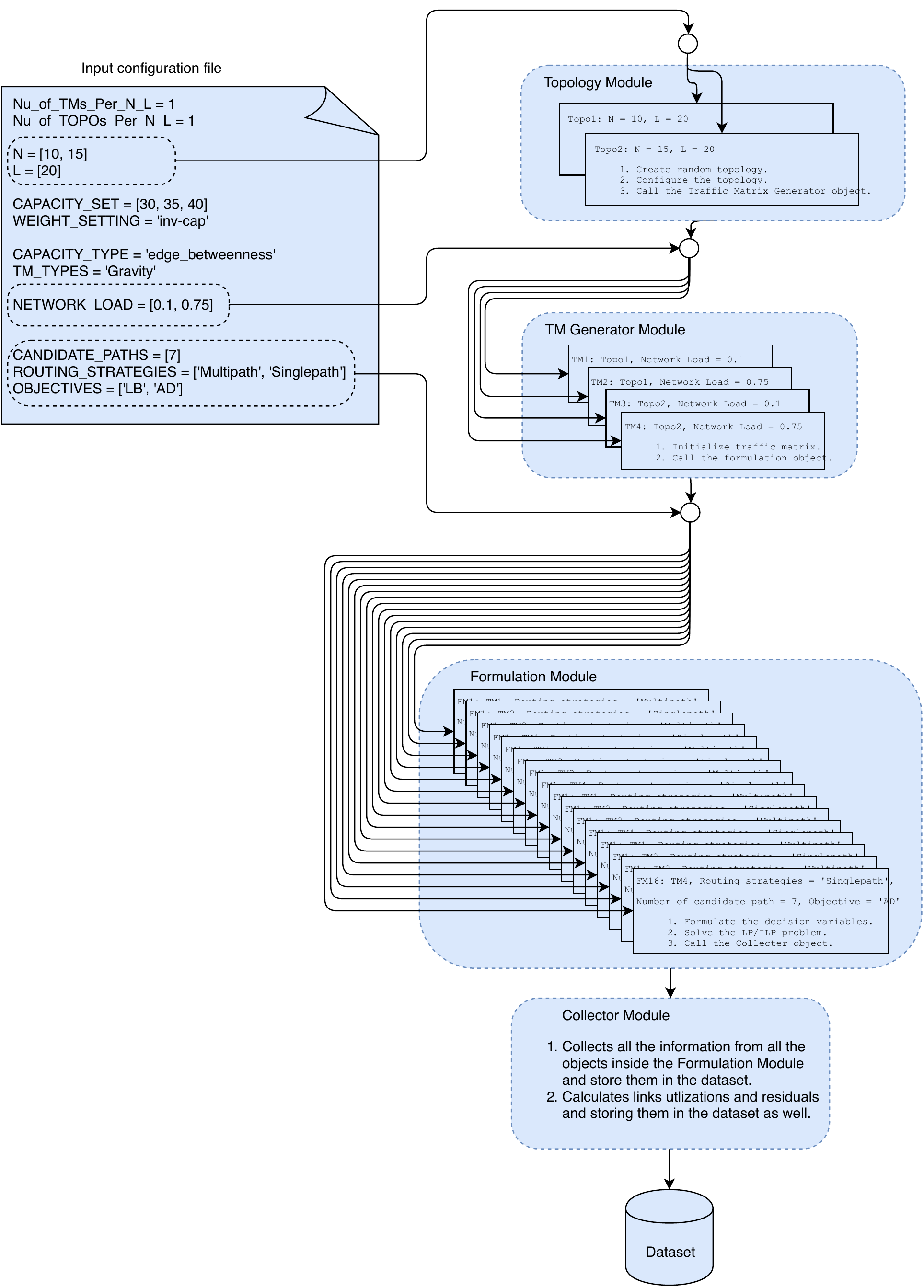}
  \caption{Design diagram showing the four main modules and an example of the input file.}
  \label{design_diagram}
\end{figure}

Efficiency is the main concern when designing this software frame for the following reasons. First, one unsolved instance of network design problem formulated as LP/MILP may have more than one million decision variables. These large number of decision variables may result from the large number of paths between any pair of nodes in the network. The number of paths grows exponentially as more nodes or links are added in the network. Second, solving an MILP problem requires more time than solving an LP problem. We tackle these problems from three aspects. First, we need to reduce the number of decision variables in the problem formulation. Second, we need to use a fast LP/MILP solver. Third, we need to come up with a design so that these formulations can be solved in parallel.
It turns out when designing a network, using all the available paths between SD pairs is unnecessary as the network tends to use shortest paths without significantly affecting optimality as explained in Section \ref{TE-Overview}. Thus, the number of decision variables can be significantly reduced when using only a few shortest paths for each SD pair. To the best of our knowledge, the fastest LP/MILP solver is Gurobi \cite{gurobi}. Compared to the default solver used in PuLP \cite{pulp}, an optimization modeler in Python, Gurobi is a lot faster.
%WWW: I changed the last sentence here because PuLp is a modeler and not a solver.
The design consists of four main modules (Figure \ref{design_diagram}). They are topology module, traffic matrix generator module, formulation module, and collector module. The topology module is responsible for instantiating objects for creating topologies and configuring them. It takes the cartesian product of the two lists N and L as input and creates a number of objects equal to the number of items in the cartesian product of these two lists. The same is true for the traffic matrix generator module and the formulation module. However, the resulting cartesian product is for each object from the previous module. In the example shown in Figure \ref{design_diagram}, each object in the topology module is composed of two objects from the traffic matrix generator module where each one is composed of four objects from the formulation module. This design pattern is like a tree structure, which makes it easier to exploit execution in parallel because each object created is independent from other objects in the same module. The traffic matrix generator module is responsible for initializing traffic matrix for all the available topologies and is based on the configuration parameters provided in the input file. The formulation module formulates all the instances as an LP/MILP and solves them after initializing the optimization model for each instance. After each instance in the formulation module is solved, an object inside the collector module is instantiated to collect all the data produced and store them in a dataset. The collector module is also responsible for some additional calculations like residual capacities and utilization of all links which are used to obtain the total residual capacity in the network.
In Figure \ref{design_diagram}, we can think of the circles as the points where the parent process spawns new child processes. These child processes are not executed all at once, but instead, they are executed in chunks according to the number of CPU cores available in the system to achieve better computational efficiency.

\section{Case Studies} \label{Data-Generation-Examples}
In this section, we present two case studies that focus on path cardinality (path budget) and routing strategies (single-path/ multipath routing). Section \ref{pathbudget1} is on path cardinality while section \ref{routing_strategies} is about routing strategies.
\begin{small}
\begin{figure*}
\centering
        \begin{subfigure}[b]{0.33\textwidth}
                \centering
                \includegraphics[trim={0 0 10mm 10mm},width=\linewidth]{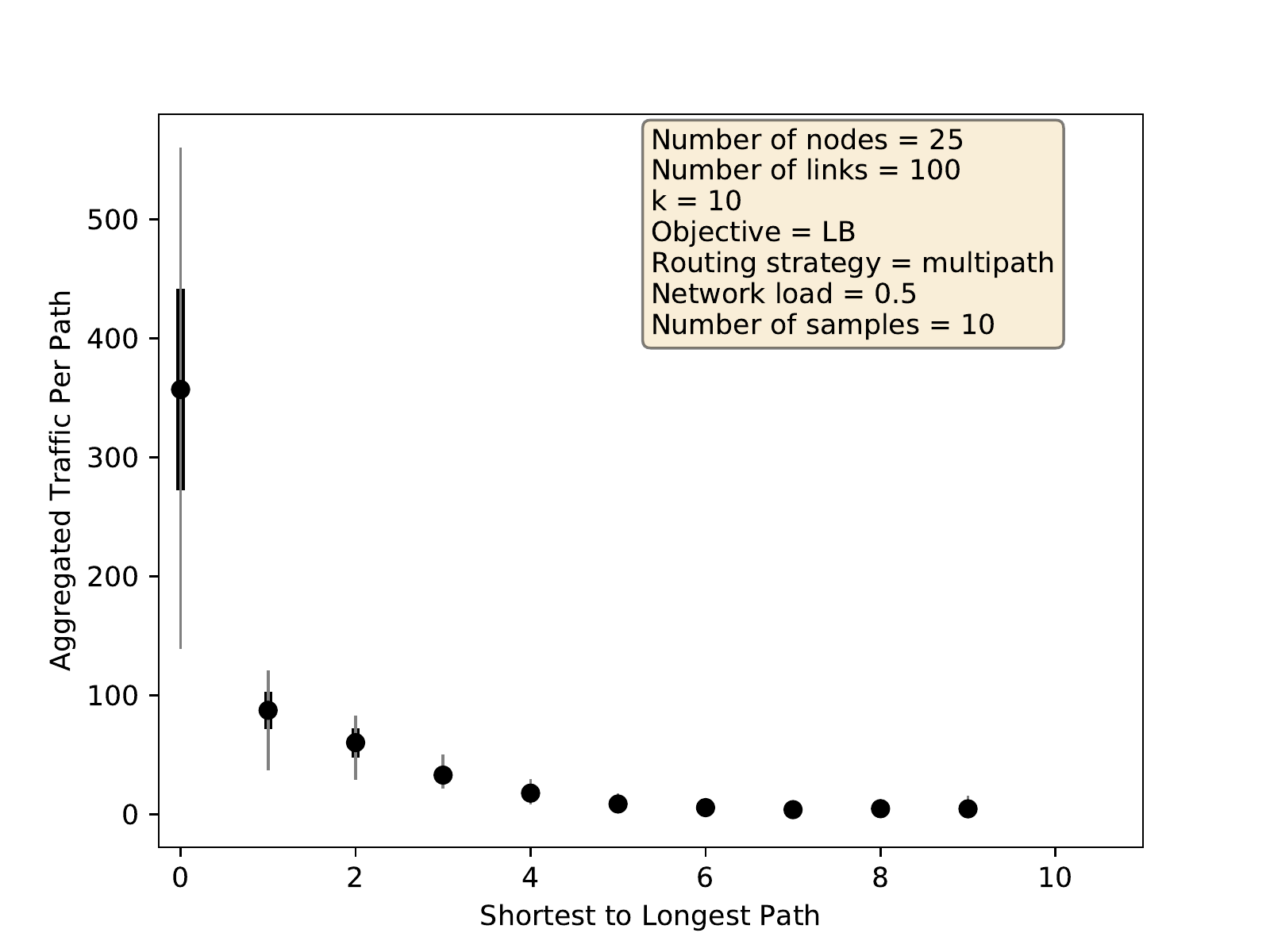}
                \caption{LB}
                \label{agg:lb}
        \end{subfigure}\hfill
        \begin{subfigure}[b]{0.33\textwidth}
                \centering
                \includegraphics[trim={0 0 10mm 10mm},width=\linewidth]{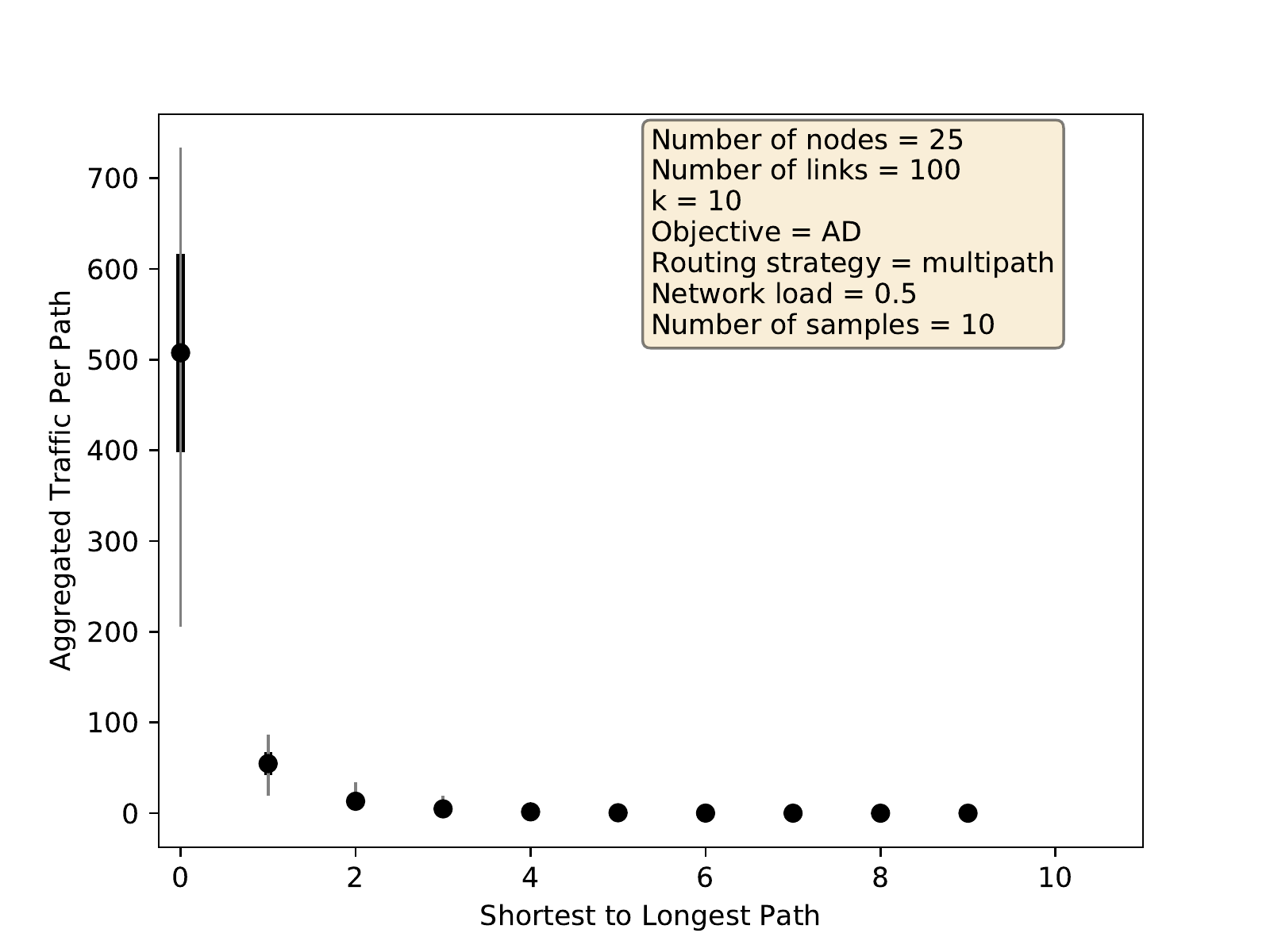}
                \caption{AD}
                \label{agg:ad}
        \end{subfigure}\hfill
        \begin{subfigure}[b]{0.33\textwidth}
                \centering
                \includegraphics[trim={0 0 10mm 10mm},width=\linewidth]{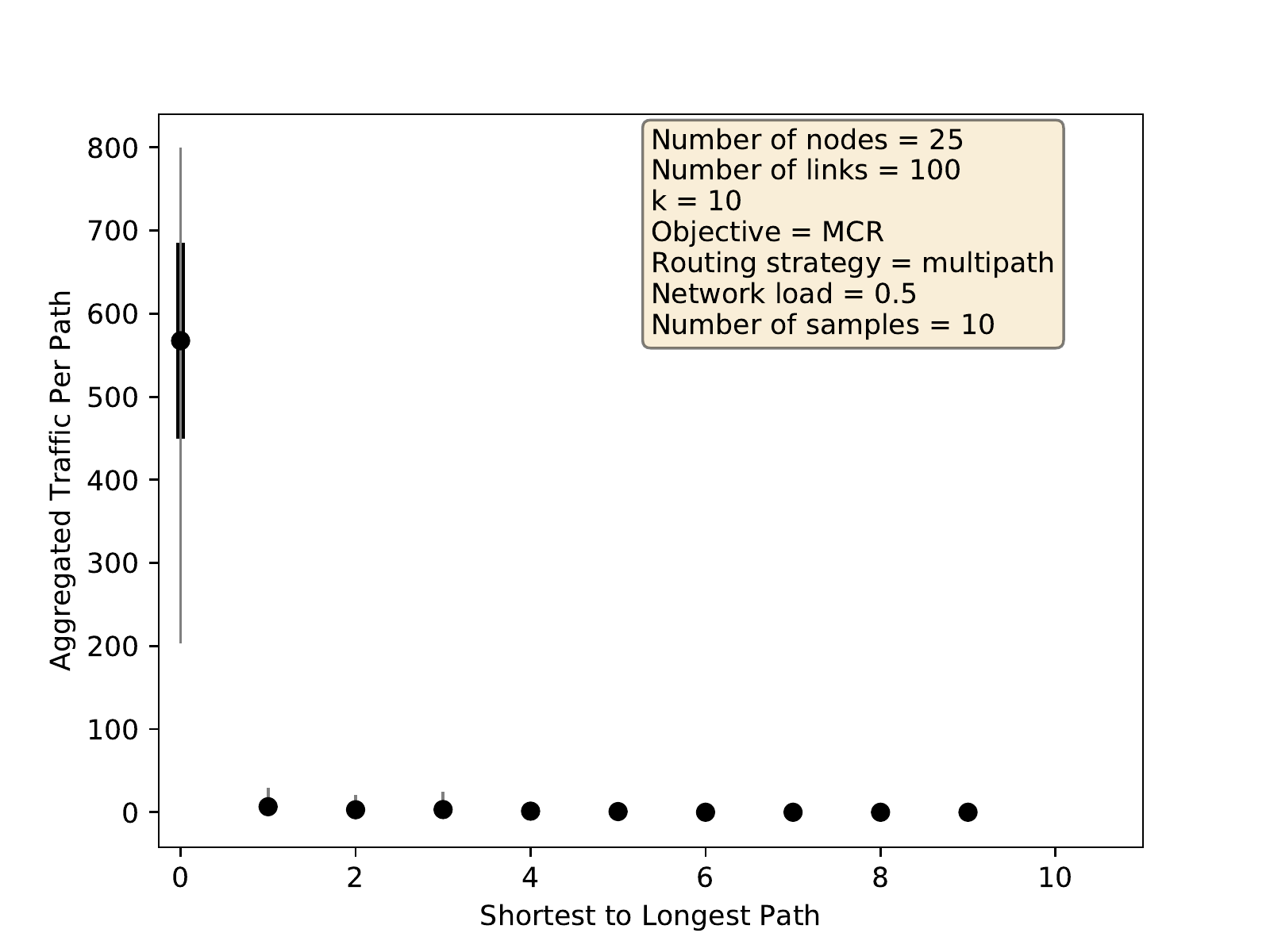}
                \caption{MCR}
                \label{agg:mcr}
        \end{subfigure}\hfill

        \caption{Aggregated traffic flow for 100 networks across ordered paths (from
the shortest to the longest).}\label{agg}
\end{figure*}
\end{small}

\subsection {Path Cardinality}\label{pathbudget1}
In this section, we give one example to illustrate a notion about path budget for networks of the same size. Then we generalize the example for networks of different sizes to depict a relationship between the nodal degree of the network and the path budget.

\subsubsection{Path Budget in Networks of Fixed Size}
The number of paths in a network grows exponentially as the size of the network increases. Thus, it is not advisable to select all of the paths in the network as this will make it very hard to solve the LP/MILP formulations in a reasonable amount of time.
Heorhiadi et. al \cite{Heorhiadi} and Ayyub et. al. \cite {SIMPLE} suggest using one of two strategies to deal with the increase of paths depending on the application, random paths for load balancing applications and shortest paths for latency-sensitive applications.
 Some researchers (\cite{peft}, \cite{deft}) were able to achieve the optimal or near-optimal solution by giving a priority to the shortest paths and directing less traffic on non-shortest paths with an exponential penalty on longer paths. Their work gave us the intuition to show that only one selection strategy may be used which is the shortest paths. We show the effectiveness of this strategy through extensive experiments and visualization of the datasets produced.
Figure \ref{agg} shows the aggregated traffic flow across ordered paths (from the shortest to the longest). In this experiment, we fix the number of nodes as (25) and the number of links as (100), and vary only the  distribution of links among nodes (traffic matrix will differ as a result). We repeat the experiment 100 times ($Nu\_of\_TOPOs\_Per\_N\_L = 100$) for each objective and then produce all the sub-figures (\ref{agg:lb}, \ref{agg:ad}, \ref{agg:mcr}). The X-axis represents the chosen number of candidate paths (ordered from the shortest path to the longest path) while Y-axis represents the aggregated traffic flow per path. The highest point in each line is the maximum aggregated traffic volume while the lowest point is the minimum aggregated traffic volume. The bold dots represent the average of the aggregated traffic volumes for each path while the bold lines represent the standard deviation. In this experiment, we consider all of the three objectives for TE (i.e., LB, AD, and MCR). Notice how the need for the next path exponentially decreases except for the MCR objective and for the most part, the first shortest path is enough to achieve superb performance.

In addition to the above observation, we also notice from these figures (\ref{agg:lb}, \ref{agg:ad}, \ref{agg:mcr}): (1) If we order the TE objectives based on the need for the next shortest path from the least to the most, we would order them in this order, i.e., MCR, AD, and LB. (2) LB uses the 2nd shortest path and other shortest paths more often than the other objectives. An intuitive explanation for this is that to achieve the LB objective, traffic must be distributed among the available paths in such a way that achieves load balancing. (3) Notice the average traffic in the first shortest path for the LB objective is below 800 which is the least among all the objectives. (4) Most of the time, MCR uses only one path which is the shortest path. An intuitive explanation is that splitting traffic would only increase the cost because the next shortest path costs more than the first shortest path (assuming that the next shortest path has a higher cost).

\begin{figure}[!t]
  \centering
   \includegraphics[width=0.45\textwidth]{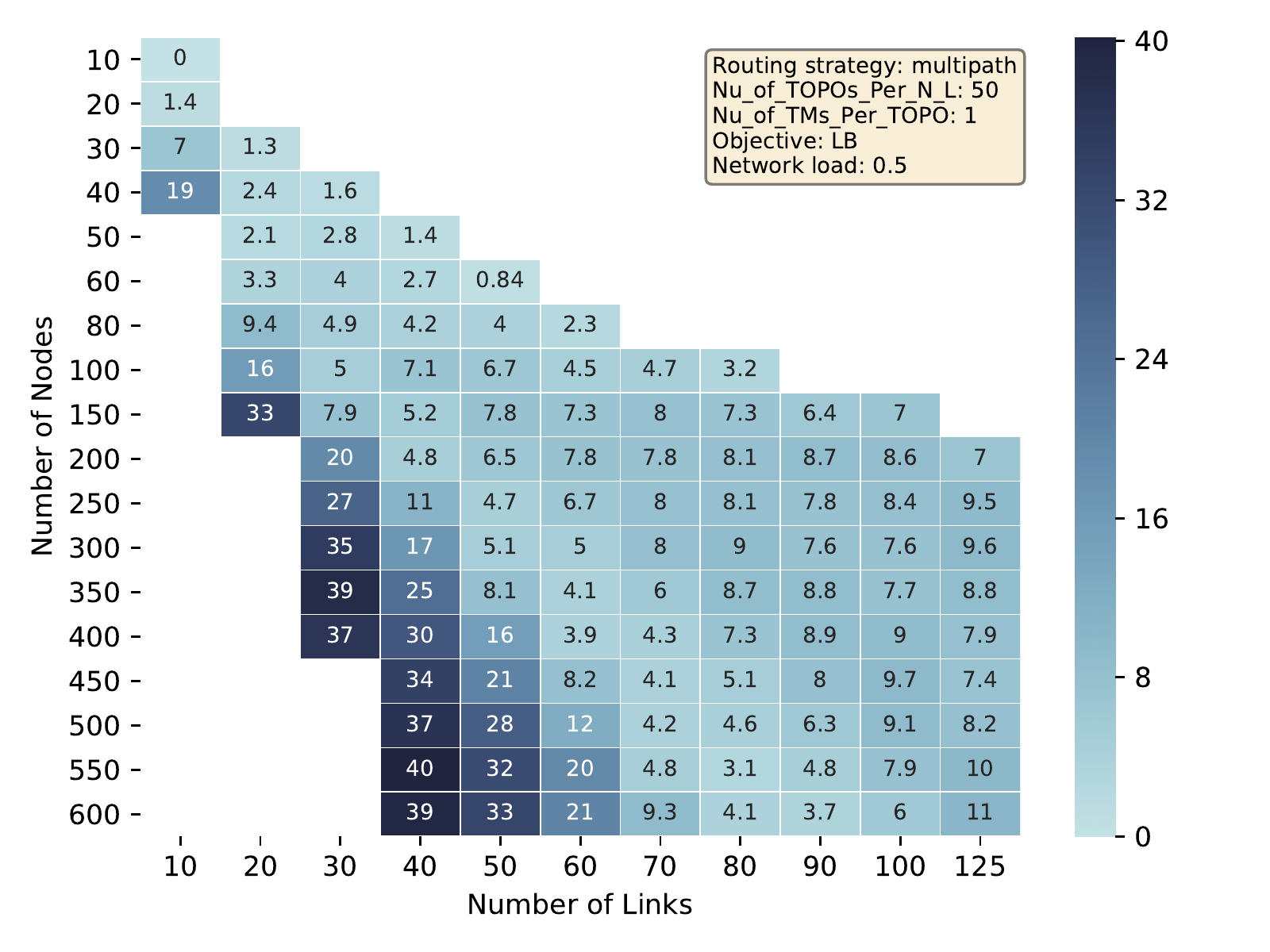}
  \caption{Average gap in LB objective for 100 instances with $k=3$ and another time with $k=7$ where $k$ is the path budget.}
  \label{deviation}
\end{figure}

\subsubsection{Path Budget in Networks of Different Sizes} \footnote{From now on we will use only the LB objective because solving LB is faster than AD. Similar to LB, AD also tends to balance load in the network. MCR is excluded because it is only used for circuit-switched network.}

%WWW: any reference for the footnote?

%WWW: I don't think there is but this was what I observed according to the simulations. LB is faster than AD because it has much less number of constraints.
The previous experiment reflects the needed number of paths to stay close to the optimal. However, we would like to know whether this range of path budget is always the same for networks of different sizes. To find out, we conduct another set of experiments with different network sizes in such a way that the topologies used reflect the relationship between the nodal degree and the required number of paths.
Thus, we present the optimality gap (Figure \ref{deviation}) that visualizes the percentage gap between two problem instances with the same settings except varying the number of candidate paths. This visualization as given in Figure \ref{deviation} (where $k$ = \{3, 7\}, the objective function is LB, $Nu\_of\_TMs\_Per\_TOPO= 100$) shows the average gap in optimality (3 paths against 7 paths) over 100 experiments for each instance. Each value represents the average gap between the objective value of the maximum $k$ ($k = 7$) and the objective value of the minimum $k$ ($k = 3$). The X-axis represents the number of nodes $N$ while the Y-axis represents the number of links $L$ (which can be thought of as the nodal degree). The gap is calculated as $((opt1-opt2)/opt1) \times 100$, where $opt1$ is the optimal value when $k=3$ and $opt2$ is the optimal value when $k=7$.
The figure made it very clear that it is not recommended to use 3 routes as candidate paths when the number of nodes is small and the number of links is large for the LB objective, i.e. when the nodal degree is large. Choosing the number of candidate paths ($k$) depends on the size of the network, nodal degree, and the objective function. Fortunately, for all the Internet Service Provider (ISP) networks that we know today, the largest nodal degree is no more than 6.0 \cite{X.Liu}. It is worth mentioning that calculating all the instances shown in Figure \ref{deviation} takes a lot of time. This is because of the large number of instances that have to be solved (one for each $k$) and all is repeated by the $Nu\_of\_TMs\_Per\_TOPO$ times. We have used one computing node with a 40-core CPU in a HPC system to produce Figure \ref{deviation} in about 8 hours.

%WWW: make sure the hardware description above is correct

%WWW: Most of the nodes at the Ohio Super Computer Center have 40 cores CPUs. Only few nodes have 80 cores which I didn't use because I am not allowed too unless I provide them a justification.

\subsection{The Effect of Multi-path and Single-path on Network Utilization} \label{routing_strategies}
Because our framework can also record link information, we study the effect of multi-path/single-path routing on link utilization. We present one example in one network and then look at performance in networks with different number of links.
We notice that there is a difference in utilization between multi-path and single-path approaches. Thus, we calculate the percentage of available capacity in the network as $R/C \times 100$ where $R$ is the summation of available residual capacities of all links and $C$ is the summation of capacities of all links.

\subsubsection{Result in One Network}
Studying the benefit of multi-path routing from only one perspective, e.g., comparing the value of the objective function, is not sufficient. For example, the LB objective, which is the minimization of the maximum congested link in the network, does not reflect the status of all links in the network. Instead, it reflects the maximum congested link and only that link. Multi-path routing still has the advantage over the single-path routing when it comes to utilization or residual capacity.

\begin{figure}[!t]
\centering
        \begin{subfigure}[t]{0.22\textwidth}
                \centering
                \includegraphics[trim={0mm 0mm 0mm 0mm},width=40mm]{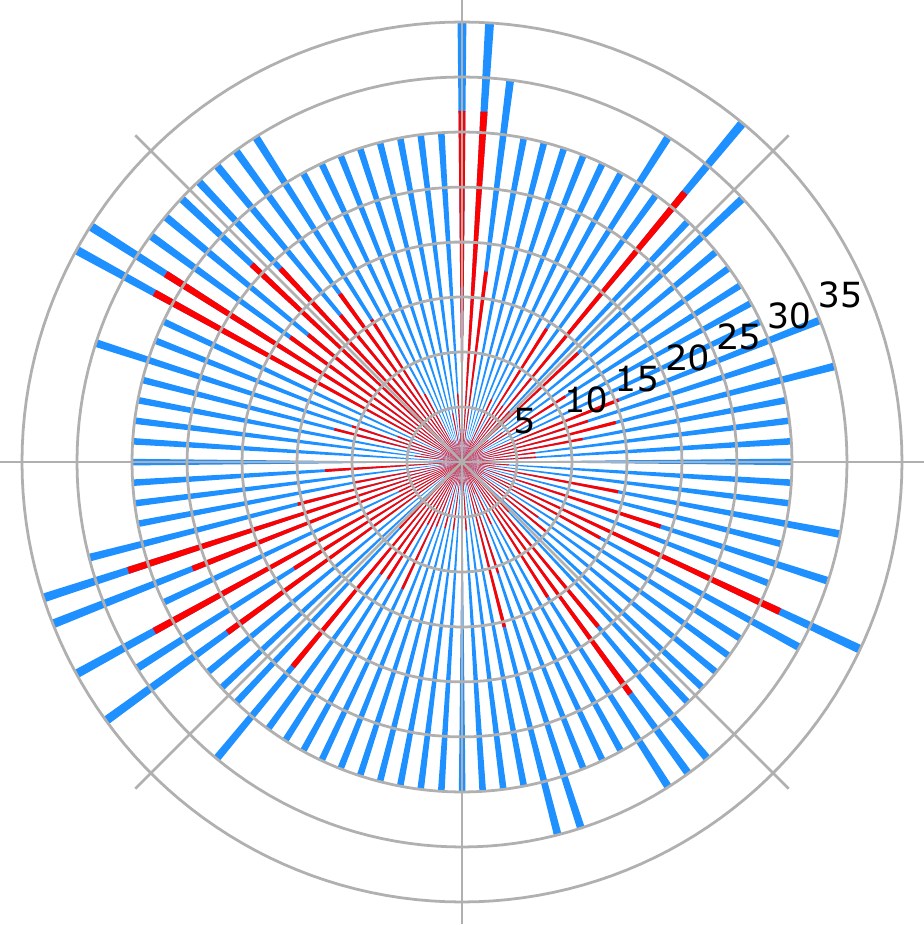}
                \caption{Multi-path case}
                \label{utilization:multi}
        \end{subfigure}\hfill
        \begin{subfigure}[t]{0.22\textwidth}
                \centering
                \includegraphics[trim={0mm 0mm 0mm 0mm}, width=40mm]{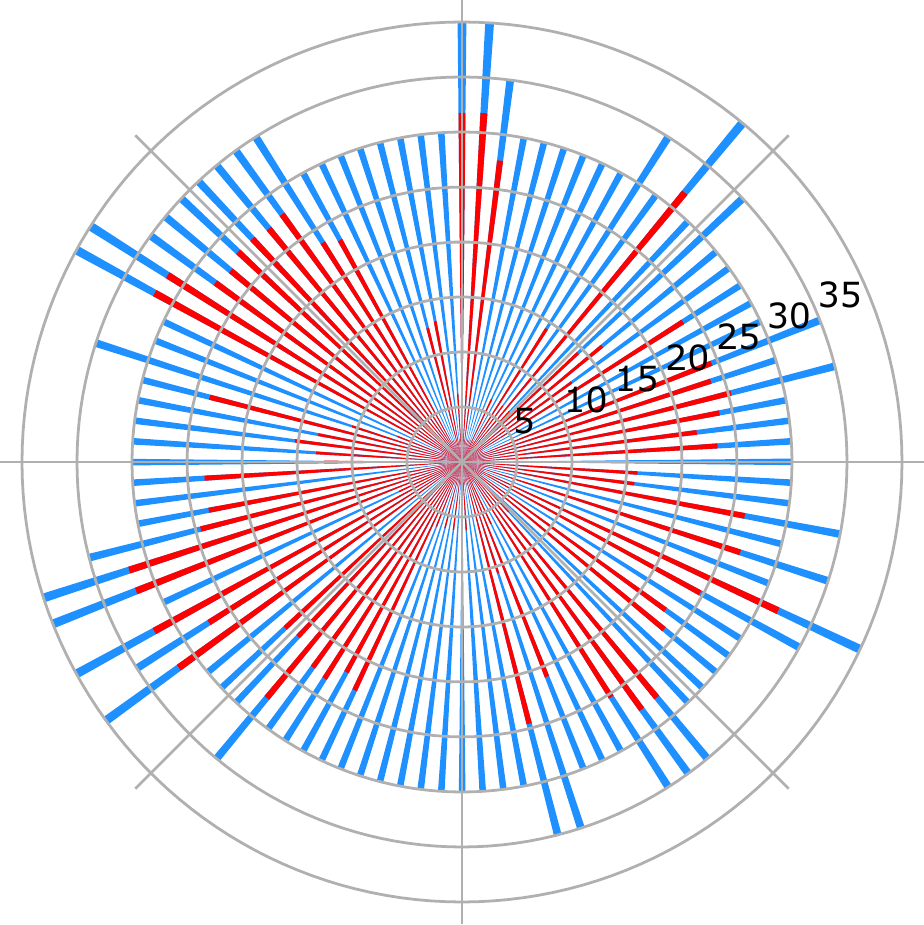}
                \caption{Single-path case}
                \label{utilization:single}
        \end{subfigure}\hfill

        \caption{A visualization that shows link load (in red) and link capacities (in blue). The single-path case utilizes the network resources 16.25\% more than the multi-path case.}\label{utilization}
\end{figure}

In this example we show how multipath/singlepath routing have an impact on the residual capacity as opposed to what was claimed in \cite{X.Liu}. It turns out, in some cases, multi-path routing utilizes links better than single-path routing. Figure \ref{utilization} shows links utilization for the multi-path case (Figure \ref{utilization:multi}) and for single-path case (Figure \ref{utilization:single}) for the same network with the same configuration. In this particular example we see that single-path routing uses more links (with higher utilization) than what multi-path routing uses, even though the value of the objective function is the same for both cases. For the multi-path routing case, the residual capacity is 67.16\% while it is 50.91\% for the single-path routing case, a difference of about 16.25\%.

\subsubsection{Result in Many Networks}
\begin{figure}[!ht]
\centering
        \begin{subfigure}[b]{0.40\textwidth}
                \centering
                \includegraphics[trim={0mm 0mm 0mm 0mm},width=60mm]{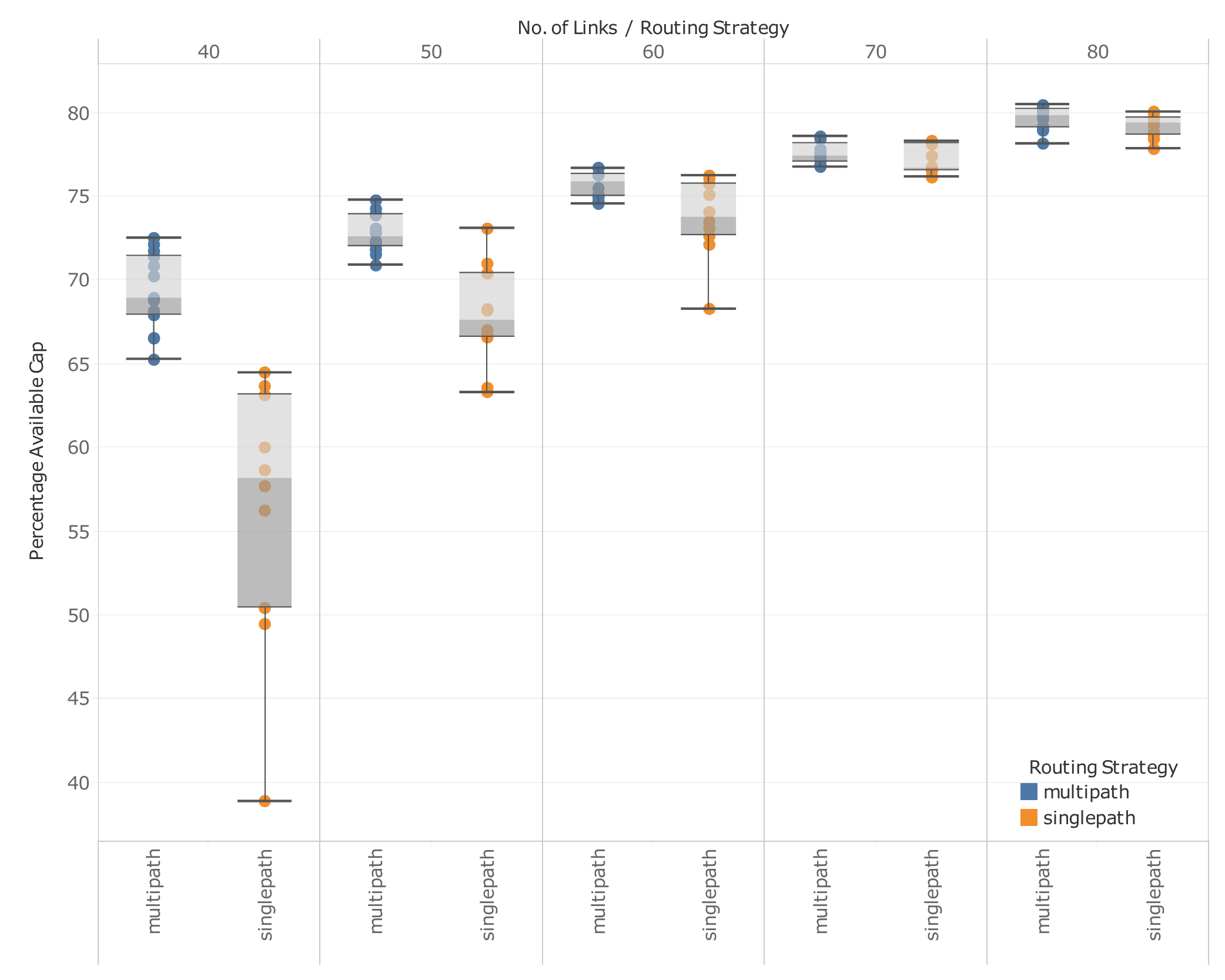}
                \caption{Available capacity \% (residual network)}
                \label{residual_cap_gap:a}
        \end{subfigure}\hfill
        \begin{subfigure}[b]{0.40\textwidth}
                \centering
                \includegraphics[trim={0mm 0mm 0mm 0mm}, width=60mm]{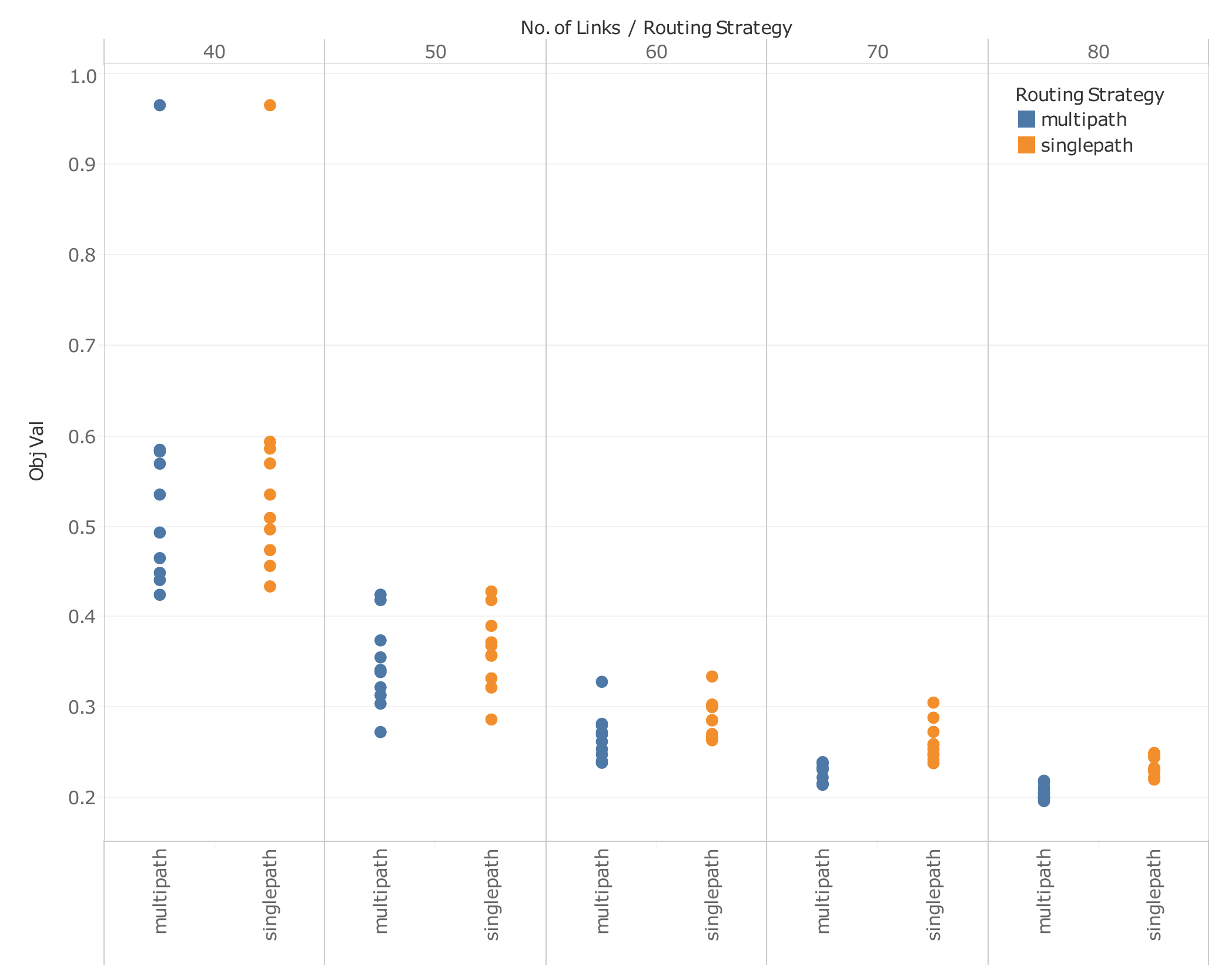}
                \caption{Value of LB objective}
                \label{residual_cap_gap:b}
        \end{subfigure}\hfill

        \caption{(a) The gap in the residual network resources between single-path and multi-path approaches. (b) shows that the LB objective stays the same for the same set of networks used in (a).}\label{residual_cap_gap}
\end{figure}

We would like to see if the observations of the last section apply to any networks.
In this example we fix the number of nodes and vary only the number of links. We see the effect of increasing the number of links on the objective function and on the residual capacity (Figure \ref{residual_cap_gap}).
We have found a relationship between the type of routing strategy (single-path/multipath) and the number of links when it comes to residual capacity. In some experiments, we notice that single-path routing over-utilizes the network resources by a significant amount. Figure \ref{residual_cap_gap:a} shows an over-utilization gap over 30\% for all topologies when the number of links is 40. This gap reduces when the number of links increases. Thus, multi-path routing is recommended when the number of links is not large. It is worth noticing that the objective value does not change that much when using multi-path and single-path routing as shown in Figure \ref{residual_cap_gap:b}.

%WWW: do not understand the caption of (b)

%WWW: I changed the caption to make it more obvious. The LB objective here represents the maximum congested link in the network.

%WWW: citation for ECMP, CSPF, use full names instead
%WWW: I have added the citations and the full names.

\section{Related Work}\label{Related-work}
Traffic engineering has been and still an active area of research for decades. The conventional approach involves adjusting links weights to find a good routing scheme that can increase throughput or minimize congestion in the network \cite{Bernard} \cite{sculpte}. It turns out that OSPF can never reach the optimal because it uses Equal Cost Multi-Path (ECMP) which splits traffic evenly among the available shortest paths \cite{peft} \cite{deft}. Furthermore, optimizing links weights is an NP-hard problem.
Recently, because of the development of software-defined networking (SDN), a centralized approach of traffic engineering came to the picture (e.g., SOL \cite{Heorhiadi}, SWAN \cite{SWAN}). At some point, all of these centralized TE systems treat the problem as an optimization problem that uses LP formulation in computing the routing scheme.
There are many packet-level simulators for computer networks (e.g., NS-3 \cite{NS-3}, OPNET \cite{OPNET} and many others). Most of these simulators model the network at a low level of abstraction. Therefore, they are not suitable for answering some of the traffic engineering questions. Moreover, flow-level simulators are much faster than packet-level simulators.  The two frameworks that are the closest to our work are YATES \cite{YATES} and REPETITA \cite{REPETITA}. However, one cannot answer our traffic engineering questions with the aid of these software frameworks. We need a framework that is easy to use and provides flexibility for configuring the network parameters and solving many instances simultaneously (i.e., software that supports the parallel execution of multi-processing).

\section{Conclusions}\label{Conclusions}
%WWW I renamed this section from (Future work and discussion) into (Conclusions)

Although our approach requires solving a large number of instances to provide insight for traffic engineering and make well-informed decision, our framework is generous, flexible, and powerful to study the effects of many different parameters in a TE system using this data-driven approach.
We have reported new results from two case studies that center on path cardinality and multi-path/single-path routing. For the first case, it turns out that the number of needed path depends on the size of the network. For the second case, we observe that in some cases multi-path routing is preferable to single-path routing for a network with small number of links to utilize the network resources efficiently.

However, there is a need to incorporate other significant TE metrics such as the stability of the TE system. This is because we may need to strike a proper balance between the optimality and network stability. Moreover, we also need to include other TE systems that do not require solving an LP/MILP problem (e.g., oblivious routing, Equal-Cost-Multi-Path (ECMP) \cite{ECMP}, Constrained-Shortest-Path-First (CSPF) \cite{CSPF}, and so on). Such TE metrics and TE algorithms will be left for future work by extending the developed framework.

\end{document}